# Design and numerical analysis of metallic Ronchi diffraction gratings acting as reflective beam-splitter

Francisco Jose Torcal-Milla[1,2,*], Luis Miguel Sanchez-Brea[2]

[1]*Grupo de Tecnología Óptica Láser, Instituto de Investigación en Ingeniería de Aragón (i3A), Universidad de Zaragoza, C/ Pedro Cerbuna 12, 50009, Zaragoza (Spain)*
[2]*Applied Optics Complutense Group, Facultad de Física, Universidad Complutense de Madrid, Plaza de las Ciencias 1, 28040, Madrid (Spain)*
[*]fjtorcal@unizar.es

## Abstract

*Objective*

In this work, a reflective beam-splitter based on a metallic Ronchi diffraction grating normally illuminated is designed and analysed. This kind of beam-splitter could have potential applications in photonics and optical technologies in which robustness is necessary since it may be manufactured over malleable metallic substrates.

*Methods*

The main idea under the design is as simple as obligating the zero-th diffraction order to be null. Firstly, scalar approach is performed, showing an approximation to the parameters of the grating necessary to achieve beam-splitting. After that, a more rigorous approach such as Rigorous Coupled Waves Analysis (TE and TM polarization) is used to evaluate the proposed diffraction gratings as reflective beam-splitters.

*Results*

Beam-splitting is demonstrated for TE and TM polarization with slightly different dimensional parameters of the diffraction grating. Besides, we show how physical height of the grating grooves that allows cancelling zero-th diffraction order for a certain illumination wavelength depends on the metals used to manufacture the grating and its period. The dependence of the grooves height on the period is exponentially decreasing. To complete the analysis, we demonstrate how for a given grating period, the grooves height also depends on the illumination wavelength.

## 1. Introduction

A diffraction grating is an optical element which periodically modulates one or some properties of the impinging light. The most common kinds of diffraction grating modulate the amplitude or the phase of the incident field [1]. In addition, gratings which modulate simultaneously the amplitude and phase [2], or the coherence state of light, have been also introduced and analyzed in recent years, [3,4]. Diffraction gratings are common optical elements used in several important fields of science, such as Chemistry, Biology, Astrophysics, Photonics, Mechanical Engineering, Robotics, etc [5,7], and also in several particular applications such as telescopes, optical encoders, machine-tool, spectroscopes, and so on [8]. The optical behavior of diffraction gratings is well known. When a plane wave illuminates the grating, Talbot effect is usually produced at the near field [9,10]. Talbot effect consists of the replication of the grating intensity pattern at distances $z_T = 2m\,p^2/\lambda$, where $p$ is the period of the grating, $\lambda$ is the illumination wavelength and $m$ are integers. The so-called self-imaging phenomenon is an interferential phenomenon and it has been analyzed with numerous configurations and conditions [11-14]. It is an useful effect in some cases for developing technology and applications [15-16] but it becomes a handicap for others such as optical encoders, whose mechanical tolerances are restricted by the Talbot effect [17-22].

On the other hand, the far field diffraction pattern of a diffraction grating perpendicularly illuminated by a plane wave consists of several diffraction orders that propagate along directions $\theta_m$ given by the grating equation, $p\sin\theta_m = m\lambda$, with $m$ integer. The power of each diffraction order depends on the profile and material which is the grating made of. Besides, the direction of the orders can be changed by changing the parameters of the grating. The simplest way consists of modifying the grating period. With more complex profiles, it is possible to obtain diffraction gratings with equal power in all orders, orders equally angularly spaced, only some orders with significant power, and so on.

In this work, a simple Ronchi grating made totally of one or two metallic substances that behaves like a beam splitter in reflection is designed and analyzed, firstly under scalar approach, and secondly by means of RCWA method, [23-25], for TE and TM polarization. The proposed grating presents relevant advantages in comparison with glass gratings such as its toughness and the possibility of manufacturing it over malleable metallic tapes, [26], which could allow curving the grating to give it a circular shape that could be useful for rotary optical encoders [27] or other applications. The proposed grating splits the beam by cancelling zero-th diffraction order in cases in which a phase Ronchi grating made of glass cannot achieve it. We reveal that the height of the grating grooves necessary to cancel zero-th diffraction order depends on the illumination wavelength, the metals used, the incident polarization, and the grating period.

Finally, we propose a possible manufacturing process by laser ablation in the case of a grating made by only one metal, or by lithographic methods in the case of a grating made of two different metals.

## 2. Scalar approach to the grating behaviour

As a first attempt of designing a reflective beam-splitter based on a metallic Ronchi diffraction grating, we analyze it under scalar approximation. Firstly, let us take a binary Ronchi lamellar phase grating made of glass, $p = 4\ \mu m$, and coated with a homogeneous and uniform chrome layer so that it acts as a phase diffraction grating in reflection configuration. This grating is illuminated perpendicularly by a monochromatic plane wave. Considering Thin Element Approximation (TEA), the reflectance of the grating can be expressed as an infinite Fourier series summation as

$$G(x) = \sum_{l=-\infty}^{\infty} a_l \exp(iqlx), \qquad (1)$$

where $a_l$ are the Fourier coefficients with $l$ integer, $q = 2\pi/p$ and $p$ is the period. Fourier coefficients depend on the physical dimensions of the grating, $(h, \tau)$, and the illumination wavelength $\lambda$, as

$$\begin{cases} a_0 = e^{i2\pi\frac{h}{\lambda}} \left\{ 1 + \tau \left[ e^{-i4\pi\frac{h}{\lambda}} - 1 \right] \right\}, & l = 0 \\ a_l = e^{i2\pi\frac{h}{\lambda}} \left( e^{-i4\pi\frac{h}{\lambda}} - 1 \right) \frac{\sin(l\pi\tau)}{l\pi}, & l \neq 0 \end{cases} \qquad (2)$$

where $\tau$ denotes the fill factor and $h$ is the thickness of the grating grooves, Figure 1. Since we have considered a Ronchi grating, then $\tau = 1/2$. We have also considered that the grating is immersed in air.

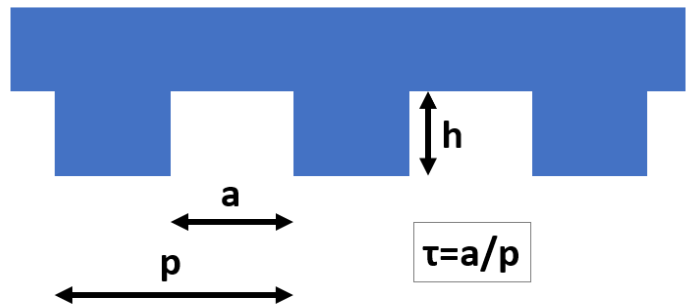

Figure 1 Scheme of the grating showing the parameters involved in the Fourier coefficients calculation.

The simplest way to achieve beam-splitting with one Ronchi grating in reflection is by imposing the zero-th diffraction order efficiency equal to zero. From Eq. (2), we obtain it by applying $h=\lambda/4$, independently on the period of the grating. Considering this, the coefficients of the other diffraction orders result

$$a_l = -i\frac{\sin(l\pi/2)}{l\pi/2}, \qquad l \neq 0. \tag{3}$$

In this case, the complex permittivity of the material is not considered and the firsts Fourier coefficients fulfill two conditions: $a_l = a_{-l}$ and even coefficients are all equal to zero.

Taking the Fresnel approximation as propagation kernel of light, the optical near field reflected by the grating when it is perpendicularly illuminated by a plane wave results

$$U(x,z) = \sum_{l=-\infty}^{\infty} a_l \exp(iqlx)\exp\left(i\frac{q^2 l^2}{2k}z\right), \tag{4}$$

where $z$ is the distance from the grating to the observation plane, $x$ is the coordinate perpendicular to the optical axis and $k=2\pi/\lambda$. The intensity is given by $I(x,z)=U(x,z)U^*(x,z)$. The reflected intensity after the grating, placed at $z=0$, under scalar approximation is shown in Figure 2a. As can be observed, quasi-continuous fringes are obtained, since the zero-th order is strictly zero. Fringes have a certain modulation along propagation because odd diffraction orders higher than first order are also present. Thus, a two-beams beam splitter cannot be achieved with this configuration and this grating period. The only way to obtain it would be by reducing the period of the grating and making higher diffraction orders evanescent.

Although, despite the scalar formalism is useful to obtain an approximation to the result, it cannot be exactly applied when we handle a real grating. Polarization of light and also complex refractive index must be taken into consideration, even more when the period of the grating approaches the illumination wavelength. Beam-splitting is still possible by cancelling zero-th diffraction order, as we will discuss along the next sections, where Rigorous Coupled Wave Analysis is used to analyze the grating design and behavior, but the thickness of the grating grooves does not result to be a quarter of the wavelength, and depends on the period of the grating, the illumination wavelength, the incident polarization, and the metals used.

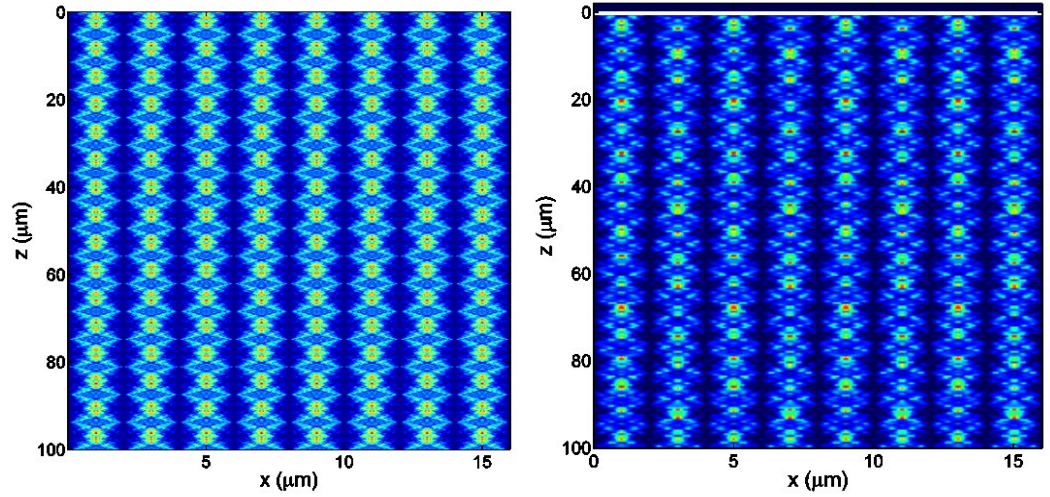

Figure 2 Near field intensity of the Chrome-coated glass grating, $p = 4\ \mu m$, $\tau = 1/2$, $\lambda = 632.8\ nm$. (a) Scalar approximation, $h = 158.2\ nm$, (b) Rigorous Coupled Wave Analysis, $h = 160\ nm$ (TE polarization).

## 3. Numerical design and analysis of the diffraction grating: rigorous approach

Let us consider a lamellar Ronchi diffraction grating acting in reflection configuration and illuminated perpendicularly to the surface by a monochromatic plane wave, λ=632.8 nm. The grating is made of chrome, whose bulk complex refractive index for the considered wavelength is $n_{Cr} = 3.4383 + 4.3405i$, [28], being $i$ the imaginary unit. A priori, its behavior will be different by considering TE or TM polarization [29,30], so we will analyze the diffraction grating behavior for both polarizations separately. Firstly, we analyze the efficiency of the lower diffraction orders in terms of the height of the grating grooves, $h$. In Figure 3 it is shown the efficiency of the zero-th to the third diffraction orders produced by the diffraction grating in terms of the height of the grooves for TE and TM polarization. Since it is still a Ronchi grating, it satisfies that $a_l = a_{-l}$. On the contrary, even orders are not strictly zero in this case, in contradiction to the scalar approach.

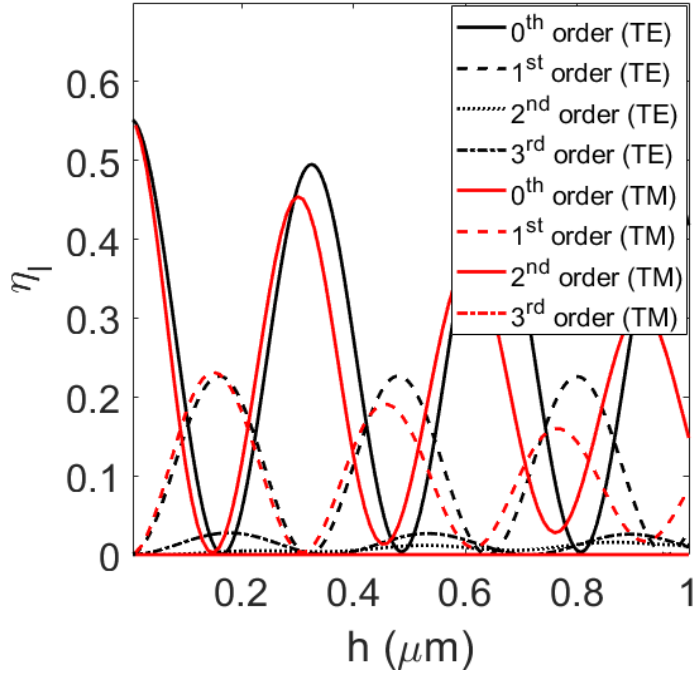

Figure 3 Efficiency of the first diffraction orders for a chrome Ronchi grating acting in reflection at normal incidence in terms of the height of the grooves obtained with RCWA (TE and TM polarization), $p = 4\ \mu m$, $\lambda = 632.8\ nm$.

Figure 3 shows that there are several values of $h$ for which zero-th order is almost zero. These heights correspond also to maxima of the first diffraction order and therefore, beam-splitting will be almost achieved for these grooves' heights. The first beam-splitting height for TE and TM polarization corresponds to $h=160$ nm and $h=153$ nm, respectively, both slightly different from the scalar prediction, that was 158.2 nm. We show in Table 1 the efficiencies of the grating diffraction orders for both polarizations calculated by using RCWA.

| $l$ | $\eta_l$ TE | $\eta_l$ TM |
|---|---|---|
| 0 | 0.0006 | 0.0003 |
| 1 | 0.2259 | 0.2205 |
| 2 | 0.0034 | 0.0016 |
| 3 | 0.0272 | 0.0221 |
| 4 | 0.0038 | 0.0018 |
| 5 | 0.0098 | 0.0082 |

Table 1. Efficiencies for the chrome grating numerically calculated using RCWA, $p = 4\ \mu m$, $h_{TE} = 160\ nm$, $h_{TM} = 153\ nm$, $\lambda = 632.8\ nm$.

For low grooves' height, only the zero-th order is present, since the grating acts like a plane mirror, Figure 3. The maximum value corresponds to the reflectance of bulked chromium at normal incidence. Consequently, other diffracted orders grow and zero-th order decreases by growing the height of the grooves. This behavior is periodical, since it is related to the phase delay produced by the grating grooves. The grating behaves almost like a mirror when the phase delay introduced by the grooves is a multiple of $2\pi$, which corresponds to different $h$ for each polarization, so it does not happen at the same instant for both cross polarizations.

In addition, it is shown in Figure 2b the near field intensity pattern produced by the mentioned chrome grating with $h$=160 nm and TE polarization. It has been considered up to the 51-th diffraction order for all simulations. As it can be observed, the fringes are pseudo-continuous. Dark fringes remain almost unalterable along the z-axis and a certain modulation is produced due to the higher diffraction orders. Besides, the period of the fringes doubles the period of the grating, as usual.

Up to this point, achieving two-beams beam-splitting is not possible. Despite of that, we may try to increase the difference between efficiencies by introducing another metal in the grating. Thus, their refractive indexes will be different and this difference will introduce an added phase delay. We have tried different combinations including chrome, gold, silver and stainless steel, being one of them the substrate and the other one forming the grooves on the substrate. In Figure 4 it is shown the efficiency of the zero-th and first diffraction orders in terms of the height of the grooves for all considered cases and both polarizations, TE and TM. Both the efficiencies and the grooves thickness to achieve minimum zero-th order efficiency are different in each case. We show in Table 2 the maximum efficiency of the first diffraction order, the minimum efficiency of the zero-th diffraction order and the grooves height necessary for all cases. With respect to higher diffraction orders, their efficiencies remain below 0.05 and we may neglect them at this point. From all tested cases, the minimum zero-th order is obtained by using the chrome grooves over stainless steel substrate for both polarizations. The bulk refractive index of stainless steel for the considered wavelength is $n_{Steel}=1.8781+3.4970i$, [31]. Again, it might be at least slightly different from values corresponding to thin layers but this approach can be taken as an approximation. On the other hand, the highest first diffraction order efficiency is obtained for the silver grooves over silver substrate grating, also for both polarizations. The bulk refractive index of silver for the considered wavelength is $n_{Ag}=0.0562+4.2760i$, [32].

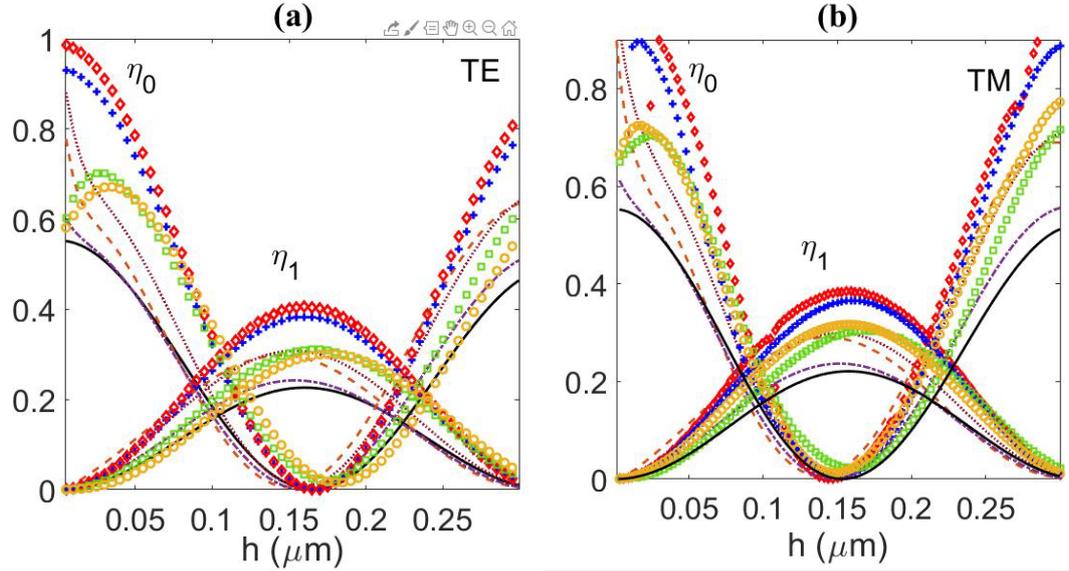

Figure 4.- Efficiency of the zero-th and first diffraction orders in terms of the thickness of the grooves for a grating with chrome grooves over chrome substrate (black solid line), chrome grooves over gold substrate (orange dashed line), chrome grooves over silver substrate (brown dotted line), chrome grooves over stainless steel substrate (purple dash-dot line), gold grooves over gold substrate (blue pluses), silver grooves over silver substrate (red diamonds), gold grooves over chrome substrate (yellow circles) and silver grooves over chrome substrate (green squares). $p = 4\ \mu m$, $\lambda = 632.8\ nm$. a) TE polarization, b) TM polarization.

*3.1 Two beams beam-splitting*

As we have mentioned before, two-beams splitting is not possible unless the period of the grating approaches the illumination wavelength, making higher diffraction orders evanescent. In this section, we achieve two beams beam-splitting (cancelling zero-th diffraction order and obtaining only $\pm 1$ diffraction orders) by using only one metallic Ronchi grating in reflection configuration. Other authors have obtained two beams splitting (sinusoidal fringes at the near field) by using much more complicated sub-wavelength gratings, mixing amplitude and phase in different levels [33] or varying the position of the grating grooves [34]. In this case, it is considered just a diffraction Ronchi lamellar metallic grating of period $p = 1\ \mu m$ illuminated with a monochromatic plane wave of wavelength λ=632.8 nm.

For clearness, we show in Figure 5 the diffraction orders efficiency for the two most interesting gratings, the whole silver grating and the chrome grooves over stainless steel substrate grating, in terms of the height of the grooves and for both polarizations. In almost all these cases, the zero-th order efficiency decays to zero and therefore perfectly continuous sinusoidal fringes are obtained, Figure 6. The fluctuations observed in Figure 6a are due to the relatively high value of the minimum zero-th order efficiency, Table 3. In addition, when the zero-th order is null, the first diffraction order achieves maximum. This means that the contrast of the fringes will be the highest possible for these heights. The first height in which the total cancellation of zero-th diffraction order occurs is shown in Table 3.

|  |  | *TE polarization* | | | *TM polarization* | | |
|---|---|---|---|---|---|---|---|
| ***Grooves*** | ***Substrate*** | *h(nm)* | $\eta_0$ | $\eta_1$ | *h(nm)* | $\eta_0$ | $\eta_1$ |
| *Chrome* | *Chrome* | 165 | 0.0006 | 0.2259 | 153 | 0.0003 | 0.2205 |
| *Gold* | *Chrome* | 150 | 0.0058 | 0.2974 | 138 | 0.0007 | 0.2901 |
| *Silver* | *Chrome* | 158 | 0.0079 | 0.3073 | 147 | 0.0096 | 0.2988 |
| *Gold* | *Silver* | 175 | 0.0001 | 0.3930 | 160 | 0.0085 | 0.3688 |
| *Silver* | *Silver* | 167 | 0.0008 | 0.4039 | 145 | 0.0008 | 0.3791 |
| *Gold* | *Gold* | 167 | 0.0004 | 0.3817 | 153 | 0.0087 | 0.3639 |
| *Silver* | *Gold* | 158 | 0.0017 | 0.3929 | 138 | 0.0025 | 0.3716 |
| *Chrome* | *Steel* | 159 | 0.0000 | 0.2420 | 147 | 0.0001 | 0.2362 |
| *Chrome* | *Silver* | 177 | 0.0099 | 0.3178 | 160 | 0.0229 | 0.3018 |
| *Chrome* | *Gold* | 168 | 0.0142 | 0.3282 | 148 | 0.0131 | 0.3139 |

Table 2. Minimum efficiency of the zero-th order, maximum efficiency of the first order and height of the grooves necessary to obtain them for some cases shown in Figure 5 and both polarizations, TE and TM. $p = 4 \ \mu m$, $\lambda = 632.8 \ nm$.

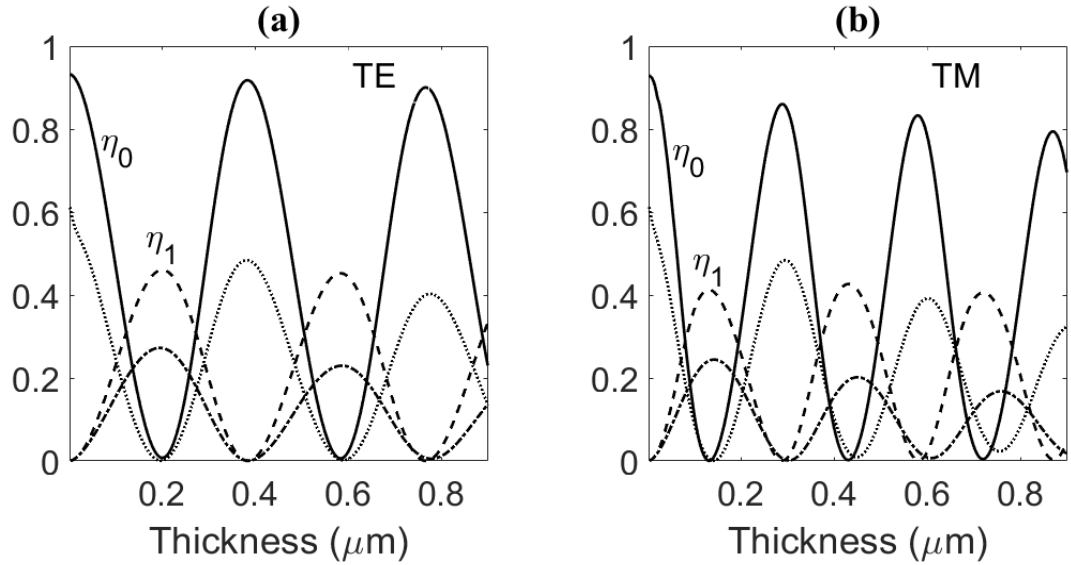

Figure 5.- Efficiency of the zero-th (solid line) and first (dashed line) diffraction orders in terms of the height of the grooves for a whole silver grating, and zero-th (dot line) and first (dash-dot line) diffraction efficiency for the grating made of chrome grooves over steel substrate. $p = 1\ \mu m$ and $\lambda = 632.8\ nm$, a) TE polarization, b) TM polarization.

|  |  | *TE polarization* |  |  | *TM polarization* |  |  |
|---|---|---|---|---|---|---|---|
| **Grooves** | **Substrate** | *h(nm)* | $\eta_0$ | $\eta_1$ | *h(nm)* | $\eta_0$ | $\eta_1$ |
| *Silver* | *Silver* | 201 | 0.0075 | 0.4604 | 132 | 0.0000 | 0.4123 |
| *Chrome* | *Steel* | 198 | 0.0005 | 0.2725 | 141 | 0.0005 | 0.2443 |

Table 3. Minimum efficiency of the zero-th order, maximum efficiency of the first order and height of the grooves necessary to obtain them for the whole silver grating and the chrome on steel grating, and both polarizations, TE and TM. $p = 1\ \mu m$, $\lambda = 632.8\ nm$.

As it is expected, a perfect quasi-sinusoidal pattern of fringes is obtained at the near field, Figure 6, since diffraction orders upper than first are evanescent and zero-th order is cancelled by manufacturing the grating with the corresponding dimensional parameters and metals. Besides, the fringes halve the period of the grating, as it was expected.

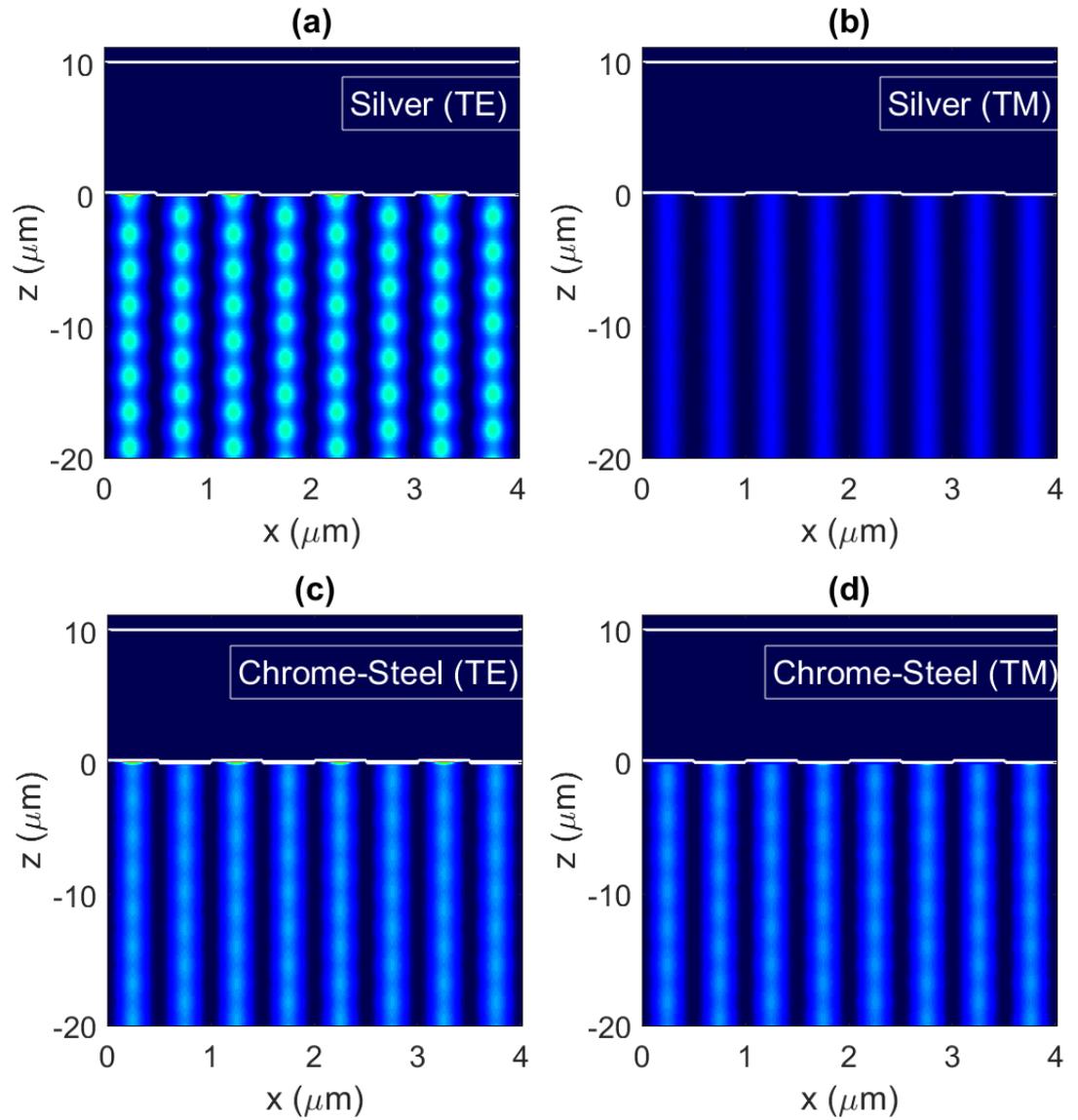

Figure 7.- Near field diffraction of the whole-silver grating at the first minimum of the zero-th order, a) TE polarization, b) TM polarization, and the chrome grooves over steel substrate grating, c) TE polarization, d) TM polarization. $p = 1\ \mu m$ and $\lambda = 632.8\ nm$.

## 3.2 Dependence on the grating period and the illuminating wavelength

It is important to notice that the thickness necessary for achieving minimum zero-th order efficiency depends on the period of the grating and the illuminating wavelength. We show in Figure 8 the dependence on the period for two diffraction gratings and a certain fixed wavelength. Both polarizations are tested, revealing similar behavior. Thickness seems to tend to a constant value for large periods, but different for TE and TM polarization. On the other hand, Figure 9 shows the dependence on the illumination wavelength for a fixed period of both diffraction gratings. Scalar approach predicts linear relationship between thickness of the grooves and wavelength, with slope 1/4. Rigorous approach also gives almost linear relationship but different for each polarization and grating period. On the other hand, for higher periods, the dependence tends to be the same for both polarizations.

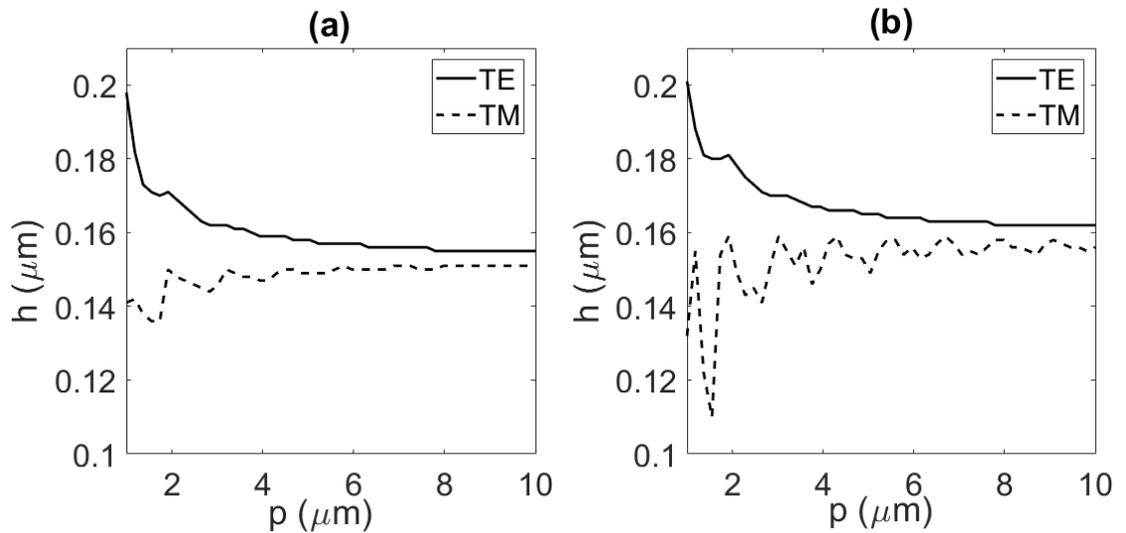

Figure 8.- Grooves thickness necessary to minimize the zero-th diffraction order efficiency in terms of the period of the grating for both polarizations. $\lambda = 632.8\ nm$. a) Chrome on steel diffraction grating, b) whole silver grating.

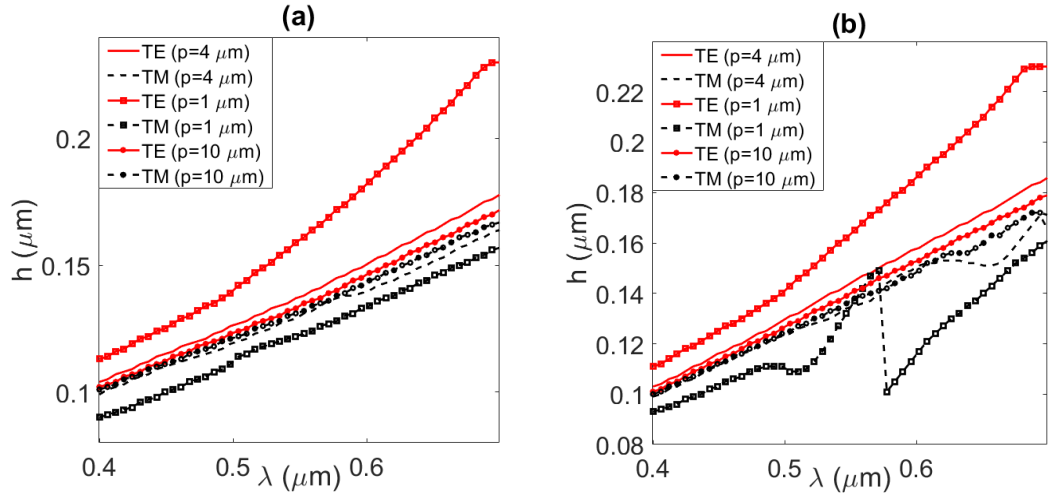

Figure 9.- Grooves thickness necessary to minimize the zero-th diffraction order efficiency in terms of the illumination wavelength for three different periods and both polarizations. a) Chrome on steel diffraction grating, b) whole silver grating.

## Conclusions

In this work, a reflective beam splitter based on a metallic lamellar Ronchi diffraction grating is design and analysed. Scalar and rigorous approaches are used to analysed the proposed gratings showing interesting results. As we demonstrate by using Rigorous Coupled Waves Analysis, the thickness of the grating grooves to achieve beam-splitting depends on the metals used, the polarization of the impinging light, the period of the grating, and the illumination wavelength. The proposed gratings can be manufactured by femtosecond laser ablation or lithographic methods, depending on the used metals. Finally, this kind of reflective beam splitter may result useful in applications in which robustness is needed, since they could be engraved over metallic substrates.

## Acknowledgements

This work has been partially supported by Gobierno de Aragón-Fondo Social Europeo-Grupo de Tecnologías Ópticas Láser (Project E44_20R), Ministerio de Ciencia e Innovación of Spain (Project PID2020-113303GB-C22), Ministerio de Economía y Competitividad (Spain) and the European funds for regional development (EU): Retos Colaboración 2019, Teluro-AEI project, RTC2019-007113-3, and by project Nanorooms PID2019-105918GB-I00.